\documentclass[twocolumn,showpacs,prd]{revtex4}
\usepackage{mathrsfs,bm,multirow}
\usepackage{longtable,lscape}
\usepackage{txfonts}
\usepackage{amssymb}
\usepackage{indentfirst}
\usepackage{graphicx,,booktabs}
\usepackage{color}
\usepackage{amssymb}

\usepackage[dvipdfm,  
            colorlinks, 
            ]{hyperref}

\usepackage{CJK}
\begin{document}
\begin{CJK}{GBK}{}
\title{The mass spectrum and strong decays of isoscalar tensor mesons}

\author{Zao-Chen Ye$^{1,2}$}\email{yezc10@lzu.edu.cn}
\author{Xiao Wang$^{1,2}$}\email{xiaowang2011@lzu.edu.cn}
\author{Xiang Liu$^{1,2}$\footnote{Corresponding author}}\email{xiangliu@lzu.edu.cn}
\affiliation{$^1$School of Physical Science and Technology, Lanzhou University, Lanzhou 730000,  China\\
$^2$Research Center for Hadron and CSR Physics,
Lanzhou University and Institute of Modern Physics of CAS, Lanzhou 730000, China}

\author{Qiang Zhao$^{1,2}$}\email{zhaoq@ihep.ac.cn}
\affiliation{$^1$Institute of High Energy Physics, Chinese Academy of Sciences, Beijing 100049, China\\
$^2$Theoretical Physics Center for Science Facilities, CAS, Beijing 100049, China}

\date{\today}

\begin{abstract}

In this work, we present a systematic study of the observed
isoscalar tensor $f_2$ states. With the detailed analysis of the
mass spectrum and calculation of the $f_2$ two-body strong decays,
we extract information of their underly structures, and try to
categorize them into the conventional tensor meson family
($n^{3}P_{2}$ $(n=1,2,3,4)$ and $m^{3}F_{2}$ ($m=1,2$) ). We also
give predictions for other decay modes of these tensor mesons, which
are useful for further experimental investigations.

\end{abstract}

\pacs{14.40.Be, 13.25.Jx, 12.38.Lg} \maketitle

\end{CJK}

\section{Introduction}\label{sec1}

During the past few decades,
there have been about 14 $f_2$ states with masses smaller than 2.5
GeV observed in experiment.  Although the Particle Data Group (PDG)
\cite{Nakamura:2010zzi} has included them in the particle list, many
of them have not yet been well-established in experimental analysis.
Taking the advantage of rich experimental information, it is
important to make a systematic analysis of these $f_2$ tensor mesons
and try to understand their properties.

In the quark model, meson typically is a bound state of a quark
($q$) and antiquark ($\bar{q}$), and $f_2$ tensor mesons have
quantum numbers $I^G(J^{PC})=0^+(2^{++})$, which means that the
relative orbital angular momentum between $q$ and $\bar{q}$ is
either $L=1$ or $L=3$, and the total spin is $S=1$. Thus, to some
extent the structure of $f_2$ tensor meson is much more complicated
comparing with that of meson with $L=0$.

Although many $f_2$ tensor states were reported by experiment, a
comprehensive understanding of their properties is still
unavailable. Firstly, how to categorizing these observed $f_2$
tensor states into the $q\bar{q}$ scenario is an intriguing question
since the total number the observed $f_2$ states is larger than that
required by the constituent quark model. It should be important to
distinguish the conventional $q\bar{q}$ tensor mesons from these
observed $f_2$ states. Secondly, a systematic theoretical study of
the decay behavior of $f_2$ tensor mesons is also absent, especially
their strong decays, which can provide abundant information for
their internal structures
\cite{Nakamura:2010zzi}.

In this work, we perform the mass spectrum analysis of the $f_2$
tensor meson family with the current experimental information. By
assigning the observed states into the quark model states with
similar masses,
we carry out the calculation of the $f_2$ two-body strong decays.
The calculated results can be compared with the experimental
measurement of the partial decay widths.


This work is organized as follows. After the introduction, we give a
brief review of the present experimental status of $f_2$ tensor
mesons. In Sec. \ref{sec3}, the mass spectrum analysis is presented.
In Sec. \ref{sec4}, the strong decays of the $f_2$ tensor mesons are
investigated in the quark pair creation (QPC) model. Finally, the
paper ends with the discussion and conclusion in Sec. \ref{sec5}.

\section{Research status of $f_2$ tensor mesons}\label{sec2}

\subsection{Experimental observations}\label{sec2exp}

The present status of the so-far observed $f_2$ states are available
in the PDG~\cite{Nakamura:2010zzi}. In Table \ref{exp}, the
resonance parameters, i.e. masses and widths, are listed.  Among
these 14 observed $f_2$ states, 6 states ($f_2(1270)$,
$f_2^\prime(1525)$, $f_2(1950)$, $f_2(2010)$, $f_2(2300)$,
$f_2(2340)$) are well established, while 6 states ($f_2(1430)$,
$f_2(1565)$, $f_2(1640)$, $f_2(1810)$, $f_2(1910)$, $f_2(2150)$) are
omitted from the summary tables of the PDG. In addition, 2 states
($f_2(1750)$, $f_2(2140)$) are listed as further states in the PDG
\cite{Nakamura:2010zzi}. As follows, we give a brief review of the
experimental status of these states.




\renewcommand{\arraystretch}{1.6}
\begin{table}[htb]
    \caption{The experimental information of $f_2$ tensor states \cite{Nakamura:2010zzi}.
    Here, the mass and the width are average values given in units of
    MeV.
    \label{exp}}
\small
\renewcommand\tabcolsep{0.08cm}
\begin{center}
    \begin{tabular}{ ccc cccccc}  \toprule[1pt]
\multicolumn{6}{c} {{\it Established states}}    \\
State&Mass&Width&State&Mass&Width\\\midrule[1pt]
$f_2(1270)$ &$1275.1\pm1.2$&$184.2^{+4.0}_{-2.4}$&$f_2^\prime(1525)$&$1525\pm5$&$73^{+6}_{-5}$ \\
$f_2(1950)$&$1944\pm 12$ &$472\pm 18$&
$f_2(2010)$&$2011^{+60}_{-80}$&$202\pm60$\\
$f_2(2300)$&$2297\pm28$&$149\pm 40$&$f_2(2340)$ &$2339\pm
60$&$319^{+80}_{-70}$\\\midrule[1pt]
\multicolumn{6}{c} {{\it States omitted from summary table of PDG}}    \\
State&Mass&Width&State&Mass&Width\\\midrule[1pt]
$f_2(1430)$&1430&&$f_2(1565)$&$1562\pm13$&$134\pm8$\\
$f_2(1640)$&$1639\pm6$&$99^{+60}_{-40}$&$f_2(1810)$&$1815\pm12$&$197\pm22$\\
$f_2(1910)$&$1903\pm9$&$196\pm31$&$f_2(2150)$&$2157\pm12$&$152\pm30$\\
\midrule[1pt]
\multicolumn{6}{c} {{\it Further states}}    \\
State&Mass&Width&State&Mass&Width\\\midrule[1pt]
$f_2(1750)$&$1755\pm10$&$67\pm12$&$f_2(2140)$&$2141\pm12$&$49\pm28$\\\bottomrule[1pt]
\end{tabular}
\end{center}
\end{table}

\subsubsection{Established states}

The signal of $f_2(1270)$ was first observed in Ref.
\cite{Boesebeck:1968zz}. Later, this state was confirmed in
reactions $\pi^{-}p\to n2\pi^{0}$ \cite{Apel:1975at,
Apel:1982ee,Alde:1998mc}, $\pi^{-}p\to n 2K^{S}_{0}$
\cite{Longacre:1986fh} and $\pi^{-}p\to 4\pi^{0}n$
\cite{Alde:1987ki}. The BESII Collaboration observed $f_2(1270)$ in
the $\pi\pi$ invariant mass spectrum of $J/\psi\to \phi
\pi^{+}\pi^{-}$ \cite{Ablikim:2004wn} and $J/\psi\to
\gamma\pi^{+}\pi^{-}$ \cite{Ablikim:2006db}.

$f^{\prime}_2(1525)$ was reported in the reactions $\pi^{-}p\to
K^{S}_{0}K^{S}_{0}n$ \cite{Crennell:1966ry,Longacre:1986fh}  and
$\pi^{-}p\to K^{+}K^{-}n$ \cite{Gorlich:1979fn,Chabaud:1981gq}.
Later, it was also confirmed by the Mark-III and BESII
collaborations in the $J/\psi$ radiative decay $J/\psi\to\gamma
K^{+}K^{-}$
\cite{Becker:1986zt,Augustin:1987fa,Bai:1996dc,Bai:2003ww}. In
addition, BESII also found $f^{^\prime}_2(1525)$ in $J/\psi\to \phi
K\bar{K}$ \cite{Ablikim:2004wn}.

$f_2(1950)$ was first reported in the  reaction
$K^{-}p\rightarrow\Lambda K\overline{K}\pi\pi$ \cite{Doser:1988fw},
and then confirmed by OMEG in $pp\rightarrow pp2(\pi^{+}\pi^{-})$
\cite{Antinori:1995wz,Barberis:1997ve}, $pp\rightarrow pp4\pi$ and
$pp\rightarrow pp2\pi2\pi^{0}$ \cite{Barberis:1999wn}. In 2000, BES
Collaboration also observed $f_2(1950)$ in $J/\psi$ radiative decay
$J/\psi\rightarrow \gamma\pi^{+}\pi^{-}\pi^{+}\pi^{-}$
\cite{Bai:1999mm}.

In 1982, a tensor structure around 2160 MeV was reported in
$\pi^-p\to \phi\phi n$ by Ref. \cite{Etkin:1982bw}. Later, the
partial wave analysis of the same reaction suggested three tensor
resonances, among which one resonance has a mass of
$2050^{+90}_{-50}$ MeV \cite{Etkin:1985se}. This signal was
confirmed by the analysis presented in Ref. \cite{Etkin:1987rj}.
This tensor structure is named as $f_2(2010)$ listed in PDG
\cite{Etkin:1985se,Etkin:1987rj}. Besides its coupling to the
$\phi\phi$ channel, $f_2(2010)$ can also decay into $KK$ and a
similar tensor structure was observed in $\pi^-p\to K_S^0K_S^0n$
with a mass of $\sim 1980$ MeV \cite{Bolonkin:1987hh} or $2005\pm
12$ MeV \cite{Vladimirsky:2006ky}.

$f_2(2300)$ was observed in $\pi^-p\to \phi\phi n$
\cite{Etkin:1982bw,Etkin:1987rj}, $\pi^-p\to K_S^0K_S^0n$
\cite{Vladimirsky:2006ky}, $\pi^-Be\to 2\phi Be$
\cite{Booth:1985kv}. In 2004, the Belle Collaboration also observed
a structure at 2.3 GeV \cite{Abe:2003vn} in $\gamma\gamma\to K^+K^-$
\cite{Abe:2003vn}, which was assigned to $f_2(2300)$.

The study of $\phi\phi$ invariant mass spectrum in $\pi^-Be\to 2\phi
Be$ reaction indicated a structure around $2392\pm 10$ MeV
corresponding to $f_2(2340)$ \cite{Booth:1985kv}. This state was
also observed in $\pi^-p\to \phi\phi n$ \cite{Etkin:1987rj} and
$p\bar{p}\to \eta\eta\pi^0$ \cite{Uman:2006xb}. Thus, the observed
decay channels of $f_2(2340)$ are $\phi\phi$ and $\eta\eta$.

\subsubsection{The $f_2$ states omitted from summary table of PDG}

$f_2(1430)$ was observed in $\pi^-p\to K_S^0 K_S^0 n$
\cite{Beusch:1967zz},  which was confirmed by the ACCMOR
Collaboration \cite{Daum:1984mj} and in Ref.
\cite{Vladimirsky:2001ek}. Later, the Axial Field Spectrometer
Collaboration found the evidence of a $2^{++}$ resonance with
$m=1480\pm50$ MeV and $\Gamma=150\pm50$ MeV in $pp\to pp\pi^+\pi^-$
\cite{Akesson:1985rn}. Although these analyses favor the same masses
around $1.43\sim 1.48$ GeV, the significant width differences
suggest that they should be different states.

$f_2(1565)$ was observed in antinucleon-nucleon annihilations and
pion-nucleon scatterings, i.e.  $\pi^-p\to \omega\omega n$
\cite{Amelin:2006wg}, $\pi^-p\to\eta\pi^+\pi^-n$
\cite{Amelin:2000nm}, $p\bar{p}\to \pi^0\eta\eta$
\cite{Amsler:2002qq}, $p\bar{p}\to \pi^0\pi^0\pi^0$
\cite{Amsler:2002qq}, $p\bar{p}\to \pi^+\pi^-\pi^0$
\cite{Bertin:1997kh,May:1990jv} and $p\bar{n}\to \pi^+\pi^+\pi^-$
\cite{Bertin:1998hu}. The observed decay modes
\cite{Nakamura:2010zzi} are $\pi\pi$, $\rho^0\rho^0$, $2\pi^+\pi^-$,
$\eta\eta$ and $\omega\omega$.

Signal for tensor state $f_2(1640)$ was first reported in $\pi^-p\to
\omega\omega n$ \cite{Alde:1988ea}. The analysis of Ref.
\cite{Adamo:1992xi} suggested a tensor structure around $1650$ MeV
in $\pi^+\pi^+\pi^-\pi^-$ in $\bar{n}p\to 3 \pi^+2\pi^-$. Bugg {\it
et al.} performed the analysis of $J/\psi\to \gamma
\pi^+\pi^+\pi^-\pi^-$, where 6 isoscalar resonances including
$f_2(1640)$ were considered for fitting the data \cite{Bugg:1995jq}.
The Crystal Barrel Collaboration carried out the partial wave
analysis of $p\bar{p}\to K^+K^-\pi^0$, where $f_2(1640)$ was also
included \cite{Amsler:2006du}.

The Bari-Bonn-CERN-Glasgow-Liverpool-Milano-Vienna Collaboration
indentified a structure around 1.8 GeV in the $K^+K^-$ system
produced in the reaction $\pi^-p\to K^+K^-n$ \cite{Costa:1980ji}. In
Ref. \cite{Cason:1982xx}, a tensor state at 1.8 GeV was observed
when performing the amplitude analysis of the reaction
$\pi^+\pi^-\to \pi^0\pi^0$. Later, the Serpukhov-Brussels-Los
Alamos-Annecy(LAPP) Collaboration studied $\pi^-p\to \eta\eta n$
\cite{Alde:1985kp}, $\pi^-p\to 4\pi^0 n$ \cite{Alde:1987ki} and
$\pi^-p\to \pi^- p 4\pi^0$ \cite{Alde:1987rn}, where a clear peak
around 1810 MeV appeared in the $4\pi^0$ mass spectrum
\cite{Alde:1987ki,Alde:1987rn}. The Belle Collaboration
\cite{Uehara:2010mq} measured $\eta\eta$ production in
$\gamma\gamma$ fusion, and found a tensor state $f_2(1810)$. In Ref.
\cite{Anisovich:2011in}, Anisovich {\it et al.} proposed that
$f_2(1810)$ was actually the same state as the $0^+$ state at 1790
MeV.

In Ref. \cite{Alde:1990qd}, $f_2(1910)$ was first observed in the
$\omega\omega$ invariant mass spectrum of $\pi^- p\to \omega\omega
n$. The WA102 Collaboration reported the evidence of $f_2(1910)$ in
$pp\to p_f(\omega\omega) p_s$ \cite{Barberis:2000kc}. In order to
describe $J^{PC}=2^{++}$ amplitudes in the $\omega\omega$ system,
$f_2(1910)$ was needed in addition to $f_2(1565)$
\cite{Amelin:2006wg}, and decay information for $f_2(1910)\to
\omega\omega$ can be extracted.

The last tensor state omitted from PDG is $f_2(2150)$. The WA102
Collaboration found the signal of $f_2(2150)$ in  $pp\to
p_f(\eta\eta)p_s$ \cite{Barberis:2000cd}, and determined the ratio
$B(f_{2}(2150)\to \eta\eta)/B(f_2(2150)\to K\bar{K})=0.78\pm0.14$.
Further experimental information of $f_2(2150)$ can be found in PDG
\cite{Nakamura:2010zzi}.

\subsubsection{Further states}

In 2006, $f_2(1750)$ was reported by analyzing $\gamma\gamma\ to
K_S^0K_S^0$ in Ref. \cite{Shchegelsky:2006et}, where the data were
collected by the L3 experiment at LEP. The resonance parameters of
$f_2(1750)$ are listed in Table \ref{exp}. The partial widths of
$f_2(1750)$ decays into $K\bar{K}$, $\pi\pi$ and $\eta\eta$ are
$17\pm 5$ MeV, $1.3\pm 1.0$ MeV and $2.0\pm0.5$ MeV, respectively
\cite{Shchegelsky:2006et}. In addition, by the $SU(3)$ analysis, the
mixing angles between nonstrange
($n\bar{n}=(u\bar{u}+d\bar{d})/\sqrt{2}$) and strange ($s\bar{s}$)
components are determined as $-1\pm 3$ degrees and $-10^{+5}_{-10}$
degrees \cite{Shchegelsky:2006et}, which correspond to two tensor
nonets $[f_2(1270),f_2^\prime(1525),a_2(1320)]$ and
$[f_2(1560),f_2(1750),a_2(1700)]$, respectively.

As a narrow enhancement, $f_2(2140)$ was observed in $\phi K^+K^-$
and $\phi\pi^+\pi^-$ final states, which are produced in $p$-$N$
interaction \cite{Green:1985pd}.

\subsection{Theoretical progress}

In the past decades,  there are many theoretical studies of the
properties of the tensor $f_2$ states. In the following, we give a
brief summary for the theoretical status of $f_2$ states.

Lattice QCD (LQCD) calculations predict that the mass of a tensor
glueball is around 2.3 GeV. It has thus initiated experimental
motivations for the search of the glueball candidate in the study of
the isoscalar tensor spectrum \cite{Chen:1994uw,Morningstar:1999rf}. Bugg
and Zou indicated that the $2^{++}$ glueball mixing with $q\bar{q}$
and $s\bar{s}$ states with $2^3P_2$ can explain why the mass of the
observed $f_2(1565)$ is lower than the expected one
\cite{Bugg:1996by}. In Ref. \cite{Barnes:1996ff},
Barnes {\it et al.} calculated the strong decays of tensor meson
with $2^3P_2$, where the mass of this state is taken as 1700 MeV.
The result shows that that $\rho\rho$, $\omega\omega$, $\pi\pi$ and
$\pi a_2$ are its important decay channels. Later, the decays of
several $f_2$ mesons with $1^3P_2$, $2^3P_2$ and $1^3F_2$ were
calculated using the $^3P_0$ model \cite{Barnes:2002mu}. By
extending the Nambu-Jona-Lasinio model, authors of Ref.
\cite{Celenza:1999ev} carried out the covariant calculation of the
properties of $f_2$ mesons. The calculated masses can be consistent
with the corresponding experimental data of $f_2$ states below 2
GeV. Here, the obtained $f_2$ states with the $s\bar{s}$ component
are around 1551 GeV and 1767 GeV, respectively, which also explains
$f_2^\prime(1525)$ and $f_2(1750)$ as 1P and 2P $s\bar{s}$ state in
the $f_2$ family, respectively \cite{Celenza:1999ev}. With the
Chiral Perturbation Theory, Dobado {\it et al.} studied the elastic
pion scattering in the $I=0$ and $J=2$ channel. They found that
$f_2(1270)$ can be described well as a pole in the second Riemann
sheet \cite{Dobado:2001rv}. Ebert {\it et al.} calculated the mass
spectra of light mesons by the relativistic quark model
\cite{Ebert:2009ub}.
Reference \cite{Surovtsev:2011hn} provides a coupled-channel
analysis of the data of $\pi\pi\to \pi\pi, K\bar{K}, \eta\eta$, and
$f_2(1270)/f_2^\prime(1525)$ and $f_2(1600)/f_2(1710)$ extracted
from the experimental data are categorized as the first and the
second tensor nonets, respectively. In Ref. \cite{Giacosa:2005bw}, the decays of the low-lying tensor mesons both into two
pseudoscalar and into pseudoscalar-vector were studied and have shown to
be in agreement with the $q\bar{q}$ interpretation. Moreover, the glueball candidate $f_J(2220)$ was also studied.

In the mixing scheme of $f_2(1270)$ and $f_2^\prime(1525)$, Li {\it
et al.}  obtained the isoscalar singlet-octet mixing angle
($\theta=27.5^\circ$) and estimated the decays of $f_2(1270)$ and
$f_2^\prime(1525)$ \cite{Li:2000zb}. Cheng and Shrock studied the
mixing of $f_2(1270)$ and $f_2^\prime(1525)$ \cite{Cheng:2011fk},
where the mixing angle of the flavor SU(3) singlet and octet was
determined as $\theta_{T,ph}=29.5^\circ$ consistent with the value
listed in PDG \cite{Nakamura:2010zzi}.

Roca and Oset suggested that $f_2(1270)$, $\rho_3(1690)$,
$f_4(2050)$,  $\rho_5(2350)$ and $f_6(2510)$ are multi-$\rho(770)$
states \cite{Roca:2010tf}. In Ref. \cite{MartinezTorres:2009uk}, the
authors studied the productions of $f_2(1270)$ and
$f_2^\prime(1525)$ via  $J/\psi\to \phi(\omega)f_2(1270)$ and
$J/\psi\to \phi(\omega)f_2^\prime(1525)$, respectively, where
$f_2(1270)$ and $f_2^\prime(1525)$ are identified as the dynamically
generated states via the vector-vector interactions from the hidden
gauge formalism \cite{Geng:2010em}. The radiative decays $J/\psi\to
\gamma f_{2}(1270)$ or $\gamma f_{2}^\prime(1525)$ were calculated
in Ref. \cite{Geng:2009iw}, and the two-photon and one photon-one
vector meson decay widths of $f_2(1270)$, $f_2^\prime(1525)$ were
obtained by treating $f_2(1270)$ and $f_2^\prime(1525)$ as the
dynamically generated states \cite{Branz:2009cv}. Ma performed a QCD
analysis for the radiative decay of heavy quarkonium with $^3S_1$
into $f_2(1270)$ and calculated the ratios of $B(\Upsilon\to \gamma
f_2(1270))$ to $B(J/\psi\to \gamma  f_2(1270))$, where the obtained
ratios are in agreement with the experimental value
\cite{Ma:2001tt}.

Besides the above theoretical work under the framework of the
conventional  meson framework, many theoretical effort have been
made in order to single out evidence for the isoscalar tensor
glueball. The glueball-$q\bar{q}$ mixing was studied using
Schwinger-type mass formulas in Ref. \cite{Burakovsky:2000gk}, which
suggested that $f_J(2220)$ should be a tensor glueball candidate. It
was also shown that $f_2(1810)$ might have a large glueball
component. In contrast, the decay of $f_2^\prime(1525)\to \pi\pi$
was consistent with its being a $q\bar{q}$ state
\cite{Burakovsky:2000gk}. The relativistic flux tube model was
applied to investigate the meson and glueball spectra, where
$f_2(1950)$ and $f_2(2010)/f_2(2300)$ can be assigned as a pure
$n\bar{n}$ and $s\bar{s}$ states, respectively, while $f_2(2340)$
was suggested as a good candidate of tensor glueball
\cite{Buisseret:2007de}. In Ref. \cite{Li:2011xr}, the mixing scheme
of $f_2(1270)$, $f_{2}^\prime(1525)$ and the $2^{++}$ glueball was
proposed.
It shows that different models have quite different prescriptions
for the classification of those observed tensor states. Because of
this, it is of great importance to provide a systematic study of the
tensor meson spectrum for any approaches. We emphasize again that
this controversial status of the isoscalar tensor spectrum motivates
us to make a systematic analysis of the strong decays of those
tensor states. More details about the experimental and theoretical
status of the tensor glueball studies can be found in Refs.
\cite{Anisovich:1990ny,Anisovich:2005iv,Crede:2008vw,Klempt:2007cp}.

\section{Mass spectrum analysis}\label{sec3}

As summarized in Sec. \ref{sec2exp},  there exist abundant
experimental observations of $f_2$ states beyond the $q\bar{q}$
scenario. As a starting point of categorizing these states, we first
try to accommodate these states into the
Regge trajectories for the mass spectrum \cite{Anisovich:2000kxa},
i.e.
\begin{eqnarray}
M^{2}=M^{2}_{0} + (n-1) \mu^{2}\label{ha} \ ,
\end{eqnarray}
where parameters $M_0$, $n$ and $\mu^2$ denote the mass of ground
state, radial quantum number and the slope parameter of the
trajectory, respectively. As shown in Ref. \cite{Anisovich:2000kxa},
$\mu^2$ is usually in the range of $1.10$ to $1.40$ GeV$^{2}$ to
give a reasonable description of  the experimental data.


\begin{center}
\begin{figure}[htb]
\begin{tabular}{c}
\scalebox{0.39}{\includegraphics{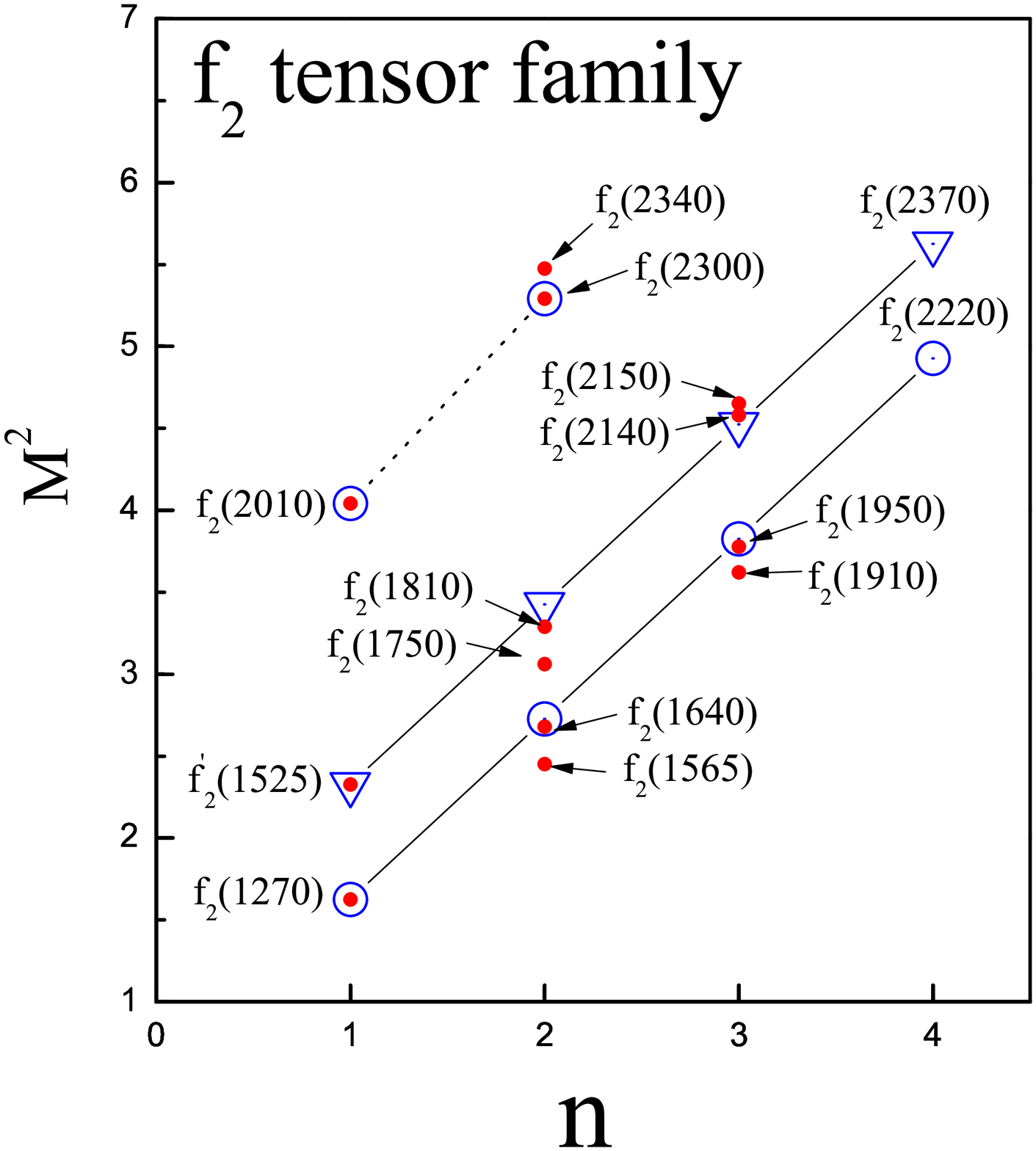}}
\end{tabular}
\caption{(color online).  The analysis of Regge trajectories for the
mass spectrum of $f_2$ family. Here, the symbols
{\color{blue}$\bigtriangledown$} and {\color{blue}$\odot$} stand for
the theoretical values of $n\bar{n}$ and $s\bar{s}$ states
respectively, which are obtained by Eq. (\ref{ha}) with $\mu^2=1.10$
GeV$^2$ and $\mu^2=1.25$ GeV$^2$ for P-wave (black solid lines) and
F-wave (black solid lines) $f_2$ mesons, respectively. While the
experimental data are marked by red dots. \label{RT}}
\end{figure}
\end{center}

As a guidance for our categorizing the $f_2$ states, we give a
comparison of the experimental data listed in Table \ref{exp} and
the result from the analysis of Regge trajectories with slope
$\mu^2=1.10$ GeV$^2$ and $\mu^2=1.25$ GeV$^2$ for P-wave and F-wave
$f_2$ mesons, respectively. Since $f_2(1270)$ and $f_2^\prime(1525)$
can be the ground states of $^{2s+1}J_L=$$^3P_2$ tensor with flavor
contents $n\bar{n}=(u\bar{u}+d\bar{d})/\sqrt{2}$ and $s\bar{s}$,
respectively, the corresponding Regge trajectories start from
$(n,M_0^2)=(1,1.270^2)$ and $(1,1.525^2)$ as shown in Fig. \ref{RT}.
One notices that the number of tensor states listed in Table
\ref{exp} is more than that of states in the $q\bar{q}$ scenario. It
indicates several possibilities: i) the signals with close masses
could be due to the same state; ii) some states cannot be
accommodated into the $q\bar{q}$ tensor meson family; iii) some
signals might be produced by artificial effects, thus, should be
omitted. Since how to distinguish these states is beyond the scope
of this work, we simply list those states with close masses in Table
\ref{RT} when comparing with the result from the analysis of Regge
trajectories.

The result shown  in Fig. \ref{RT} indicates that
$[f_2(1565)/f_2(1640)$, $f_2(1750)/f_2(1810)]$,
$[f_2(1910)/f_2(1950)$, $f_2(2120)/f_2(2140)/f_2(2150)]$ and
$[f_2(2220), f_2(2370)]$ can be organized as the first, second, and
third radial excitations of $[f_2(1270), f_2^\prime(1525)]$,
respectively. As discussed above, we cannot distinguish $f_2(1565)$
and $f_2(1640)$  by the analysis of Regge trajectories since their
masses are close to each other. The situations for
$f_2(1750)/f_2(1810)$, $f_2(1910)/f_2(1950)$ and
$f_2(2140)/f_2(2150)$ are similar to that of $f_2(1565)/f_2(1640)$.
Thus, further study of strong decay behavior of these $f_2$ states
will be helpful to clarify their properties. The details will be
given in Sec. \ref{sec4}.

We also try to group those $P$-wave $f_2$ mesons in association with
the corresponding $a_2$ and $K_2$ mesons to form tensor nonets, i.e.
\begin{eqnarray*}
&&1P: f_2(1270), f_2^\prime(1525), a_2(1320), K_2^*(1430), \\
&&2P: \left\{\begin{array}{c} f_2(1565)\\f_2(1640) \end{array}
\right., \left\{\begin{array}{c} f_2(1750)\\f_2(1810) \end{array}
\right., a_2(1700),  K_2^*(?),\\
&&3P: \left\{\begin{array}{c} f_2(1910)\\f_2(1950) \end{array}
\right., \left\{\begin{array}{c} f_2(2140)\\f_2(2150) \end{array}
\right., a_2(1950), K_2^*(1980),\\
&&4P: f_2(2220), f_2(2370), a_2(2250), K_2^*(?),
\end{eqnarray*}
where two tensor $K_2^*(?)$ mesons corresponding to $2P$ and $4P$
states are absent in experiment.

The mass spectrum analysis shown in Fig. \ref{RT} also indicates
that $f_2(2010)$ and $f_2(2300)$ are possible candidates for the
$1^3F_2$ and $2^3F_2$ states, respectively, when taking the slope
$\mu^2=1.25$ GeV$^2$.
In the next section, we will discuss the possibility of treating the
reported $f_2(2300)$  or $f_2(2340)$ as the first radial excitation
of $f_2(2010)$.

\section{Strong decay behavior}\label{sec4}

For obtaining the two-body strong decay behavior of these  discussed
$f_2$ states, we adopt the Quark Pair Creation (QPC) model
\cite{Micu:1968mk}, which have been extensively applied to the study
of the strong decay of hadrons
\cite{yaouanc,LeYaouanc:1977gm,LeYaouanc:1988fx,vanBeveren:1979bd,vanBeveren:1982qb,Bonnaz:2001aj,sb,Lu:2006ry,Luo:2009wu,
Blundell:1995ev,Page:1995rh,Capstick:1986bm,Capstick:1993kb,Ackleh:1996yt,Close:2005se,Zhou:2004mw,Guo:2005cs,Zhang:2006yj,Chen:2007xf,Li:2008mz,Sun:2009tg,Liu:2009fe,Sun:2010pg,Yu:2011ta,Wang:2012wa}.
For the calculation of two-body strong decay of a hadron, the
operator $T$ accounting for the $q\bar{q}$ creation from the vacuum
is introduced by
\begin{eqnarray}
T&=& - 3 \gamma \sum_m\: \langle 1\;m;1\;-m|0\;0 \rangle\, \int {\rm
d}{\bm{k}}_3 \; {\rm d}{\bm{k}}_4 \delta^3({\bm{k}}_3+{\bm{k}}_4)
\nonumber\\&&\times{\cal
Y}_{1m}\left(\frac{{\bm{k}}_3-{\bm{k}_4}}{2}\right)\; \chi^{3 4}_{1,
-\!m}\; \varphi^{3 4}_0\;\, \omega^{3 4}_0\;
d^\dagger_{3i}({\bm{k}}_3)\;
b^\dagger_{4j}({\bm{k}}_4)\,\label{tmatrix}
\end{eqnarray}
where the definitions of flavor singlet, color singlet and the
$\ell$th solid harmonic polynomial are as follows
\begin{eqnarray*}
\varphi^{34}_{0}&=&\frac{u\bar u +d\bar d +s \bar
s}{\sqrt 3},\,\,
\omega_{0}^{34}=\frac{1}{\sqrt{3}}\delta_{\alpha_3\alpha_4}\,(\alpha=1,2,3),\\
&&\qquad\mathcal{Y}_{\ell m}(\bm{k})=
|\bm{k}|^{\ell}Y_{\ell m}(\theta_{k},\phi_{k}).
\end{eqnarray*}
In Eq. (\ref{tmatrix}), $i$ and $j$ are the $SU(3)$ color indices of
the created quark and anti-quark, and $\chi_{{1,-m}}^{34}$ denotes a
spin triplet state.

The transition matrix element for $A$ decay into $B$ and $C$ can be
expressed in terms of the helicity amplitude as
\begin{eqnarray}
\langle{}BC|T|A\rangle=\delta^3(\bm{K}_B+\bm{K}_C-\bm{K}_A)\mathcal{M}^{M_{J_A}M_{J_B}M_{J_C}}.
\end{eqnarray}
By the Jacob-Wick formula \cite{Jacob:1959at},  the partial wave
amplitude $\mathcal{M}^{J L}$ can be further related to the helicity
amplitude $\mathcal{M}^{M_{J_A} M_{J_B} M_{J_C}}$. Thus, the partial
decay width can be written as
\begin{eqnarray}
\Gamma = \pi^2 \frac{{|\bm{K}|}}{M_A^2}\sum_{JL}\Big
|\mathcal{M}^{J L}\Big|^2,
\end{eqnarray}
where $|\bm{K}|$ is the three momentum of the daughter hadrons in
the initial state center of mass (c.m.) frame.

In Eq. (\ref{tmatrix}), a dimensionless parameter $\gamma$ is
introduced for describing the strength of the quark pair creation
from the vacuum. It can be extracted by fitting the experimental
data, for which 15 decay channels are included as listed in Table
\ref{fit}. In the numerical calculation of the partial decay width,
we adopt the harmonic oscillator wave function for the spatial wave
function of the meson, {\it i.e.}, $\Psi_{n,\ell
m}(R,\bm{k})=\mathcal{R}_{n,\ell}(R,{\bm k})\mathcal{Y}_{n,\ell
m}({\bm k})$, where ${R}$ is determined by reproducing the
realistic root mean square radius by solving the Schr\"{o}dinger
equation with the linear potential \cite{Close:2005se,Yu:2011ta}, i.e., we have relation
$$\sqrt{\langle r^2 \rangle}\equiv\bigg(\int \Psi^*_{n,\ell
m}(R,\bm{r}) r^2 \Psi_{n,\ell
m}(R,\bm{r}) d^3r\bigg)^{1/2},$$
where $\sqrt{\langle r^2 \rangle}$ is from solving the Schr\"{o}dinger
equation with the linear potential. By this relation, we finally get the $R$ value for the corresponding meson.
The resonance parameters of the mesons involved in our calculation
are taken from the data listed in PDG \cite{Nakamura:2010zzi}.

We define
$\chi^{2}=\sum_{i}(\Gamma_{i}^{theory}-\Gamma^{exp}_{i})^{2}/\delta^{2}_{\Gamma_{i}}$,
where $\delta_{\Gamma_i}$ denotes the average experimental error of
each partial decay width. By minimizing $\chi^{2}$, we obtain
$\gamma_{0}=8.7$, and the corresponding experimental data and
theoretical results are shown in Table \ref{fit}.

In the following,  we present the numerical results for the partial
decay widths of those isoscalar tensor states.

\begin{table}[htb]
\centering%
\caption{The measured partial decay widths of 16 decay channels
and the comparison with theoretical calculation (the third
column). Here, the minimum of $\chi^2$ is $2149$.   \label{fit}}
\begin{tabular}{lccccccc}
\toprule[1pt]
   Decay channel&                                    Measured width (MeV) \cite{Nakamura:2010zzi} &        QPC (MeV) \\\midrule[1pt]
$b_{1}(1235)\rightarrow \omega\pi$&            142$\pm$8 &        119.5        \\
$\phi\rightarrow K^{+}K^{-}$&                2.08$\pm$0.02 &      1.82           \\
$a_{2}(1320)\rightarrow \eta\pi$&            15.5$\pm$0.7 &       22.6           \\
$a_2(1320)\rightarrow K\overline{K}$&        5.2$\pm$0.2 &         2.1          \\
$\pi_2(1670)\rightarrow f_2(1270)\pi$&       145.8$\pm$5.1 &    118.7             \\
$\pi_2(1670)\rightarrow \rho\pi$&            80.3$\pm$2.8 &      70.1              \\
$\rho_{3}(1690)\rightarrow \pi\pi$&           38$\pm$2.4 &      57.9              \\
$\rho_{3}(1690)\rightarrow \omega\pi$&        25.8$\pm$1.6 &    71.7         \\
$\rho_{3}(1690)\rightarrow K\overline{K}$&        2.5$\pm$0.2 &     1.3           \\
$K^{*}(892)\rightarrow K\pi$&                48.7$\pm$0.8 &        28.4           \\
$K^{*}(1410)\rightarrow K\pi$&                15.3$\pm$1.4 &      13.3         \\
$K^{*}_{0}(1430)\rightarrow K\pi$&                251$\pm$74 &    165.5         \\
$K^{*}_{2}(1430)\rightarrow K\pi$&             54.4$\pm$2.5 &    66.8                \\
$K^{*}_{2}(1430)\rightarrow K^{*}(892)\pi$&    26.9$\pm$1.2 &      33.6            \\
$K^{*}_{2}(1430)\rightarrow K\rho$&    9.5$\pm$0.4&      13.2            \\
$K^{*}_{2}(1430)\rightarrow K\omega$&    3.16$\pm$0.15 &      3.9            \\
\bottomrule[1pt]
\end{tabular}
\end{table}

\subsection{The ground states in tensor meson family}

As the candidate of $1^{3}P_{2}$ tensor states, $f_2(1270)$ and
$f^\prime_2(1525)$ can be regarded as the mixture of
$N=(u\bar{u}+d\bar{d})/{\sqrt{2}}$ and $S=s\bar{s}$
\begin{eqnarray}
|f_2(1270)\rangle&=&\sin\phi|N\rangle+\cos\phi|S\rangle,\label{1}
\\
|f^\prime_2(1525)\rangle&=&\cos\phi|N\rangle-\sin\phi|S\rangle\label{2}
\end{eqnarray}
with $\phi\equiv \theta+54.7^\circ$,  where $\theta$ is the mixing
angle between the SU(3) flavor singlet and octet. In our
calculation, we take $\theta=29.6^\circ$ from PDG
\cite{Nakamura:2010zzi}, which corresponds to $\phi=84.3^\circ$. In
Ref. \cite{Giacosa:2005bw}, a similar value for the mixing angle
$\theta$ was found.


\begin{center}
\begin{figure}[htb]
\begin{tabular}{c}
\includegraphics[bb=152 305 460 730,scale=0.8,clip]{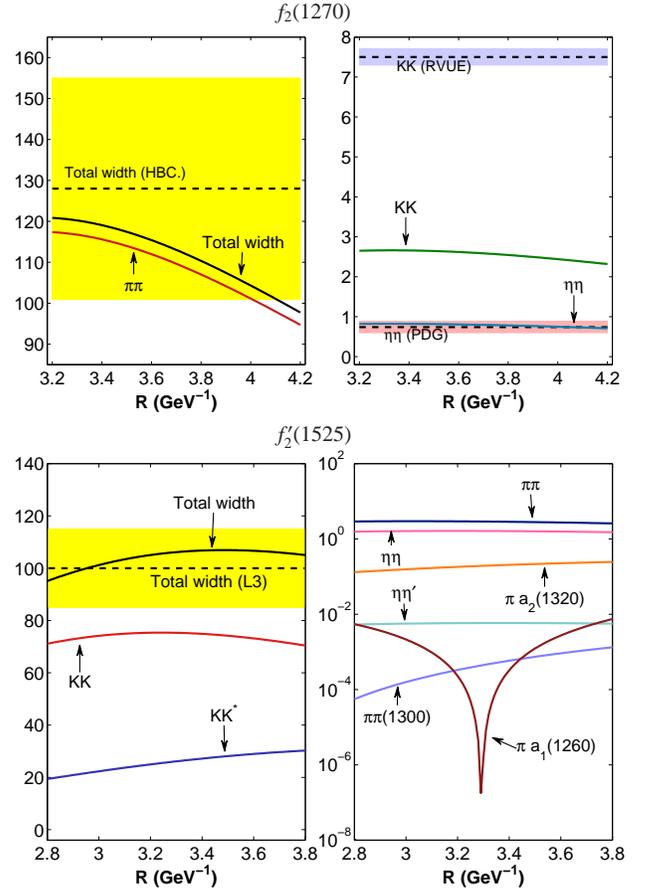}
\end{tabular}
\caption{(color online).  The obtained two-body partial decay widths
of $f_2(1270)$ and $f_2^\prime(1525)$ dependent on the $R$ value and
the comparison with the experimental data (the dashed lines with
bands). Here, the experimental width of $f_2(1270)$ and
$f_2^\prime(1525)$ are taken from Refs. \cite{Boesebeck:1968zz} and
\cite{Acciarri:2000ex}, respectively. The experimental partial
widths of $f_2(1270)$ decays into $K\bar{K}$ and $\eta\eta$ are
given by Refs. \cite{Shchegelsky:2006et} and
\cite{Nakamura:2010zzi}, respectively. \label{1270/1525}}
\end{figure}
\end{center}

In Fig. \ref{1270/1525},  we present the calculated partial decay
widths of $f_2(1270)$ and $f_2^\prime(1525)$ in terms of parameter
$R$ 
within a typical range of values. For $f_2(1270)$, our calculation
indicates that $\pi\pi$ channel is its dominant decay mode and
$\Gamma(f_2(1270)\to K\bar{K})>\Gamma(f_2(1270)\to \eta\eta)$. These
decay behaviors are consistent with the experimental observation
\cite{Nakamura:2010zzi}. We also notice that the PDG data
($\Gamma_{\eta\eta}=0.74\pm0.14$ MeV) \cite{Nakamura:2010zzi} can be
described by our calculation of the decay width of $f_2(1270)\to
\eta\eta$ well. However, the obtained $\Gamma(f_2(1270)\to
K\bar{K})$ is about $0.27\sim0.35$ times smaller than the data
\cite{Nakamura:2010zzi}.

The sum of the theoretical two-body strong decays of  $f_2(1270)$
can reach up to $120.5\sim 97.5$ MeV corresponding to $R=3.2\sim
4.2$ GeV$^{-1}$. This value is smaller than the experimental average
width $\Gamma_{f_2(1270)}=184.2^{+4.0}_{-2.4}$ MeV
\cite{Nakamura:2010zzi}, but comparable with the results of
measurements of Refs. \cite{Boesebeck:1968zz,Tikhomirov:2003gg}.
Actually, there still exist large experimental discrepancies among
different experimental measurements of the $f_2(1270)$ width as
listed in PDG \cite{Nakamura:2010zzi}. Further experimental study of
the resonance parameter of $f_2(1270)$ is still needed. In addition,
we note that we do not include the partial width of $f_2(1270)$
decays into multipions (the sum of the branching ratios of
$f_2(1270)\to \pi^+\pi^-2\pi^0, 2\pi^+2\pi^-$ is about $9.9\%$
\cite{Nakamura:2010zzi}) when making the comparison between our
calculation and the experimental data. Thus, the difference between
our result and the central value of the $f_2(1270)$ width
\cite{Boesebeck:1968zz} shown in Fig. \ref{1270/1525} can be
understood.

The results for $f_2^\prime(1525)$ is presented in the lower panel
of Fig. \ref{1270/1525}. The sum of the two-body strong decays of
$f_2^\prime(1525)$ calculated in this work overlaps with the
experimental data
from seven experiments, i.e. $108^{+5}_{-2}$ MeV
\cite{Longacre:1986fh}, $102\pm42$ MeV \cite{Tikhomirov:2003gg},
$100\pm15$ MeV \cite{Acciarri:2000ex},  $90\pm12$ MeV
\cite{Aston:1987am}, $103\pm30$ MeV \cite{Augustin:1987fa},
$104\pm10$ MeV \cite{Shchegelsky:2006et}, $100\pm3$ MeV
\cite{Falvard:1988fc}. The calculated $\Gamma(f'_2(1525)\to
K\bar{K})$ is about $70.4\sim75.3$ MeV corresponding to
$R=2.8\sim3.8$ GeV$^{-1}$, which is comparable with the experimental
average value $65^{+5}_{-4}$ MeV listed in PDG
\cite{Nakamura:2010zzi}. Our result suggests
$\Gamma(f^\prime_2(1525)\to K\bar{K}^{*}+h.c.)$ is about 24.1 MeV
($R=3.125$ GeV$^{-1}$), which shows that $K\bar{K}^{*}+h.c.$ is an
important decay channels of $f_2^\prime(1525)$. We also suggest
future experiment to carry out the search for $f_2^\prime(1525)$ in
the $K\bar{K}^{*}+h.c.$ decay channel. We obtain $\Gamma(
f^\prime_2(1525)\to \eta\eta)=1.63$ MeV with the typical value
$R=3.125$ GeV$^{-1}$, which is smaller than the experimental data
($\Gamma_{\eta\eta}=5\pm0.8$ MeV \cite{Shchegelsky:2006et}). The
obtained $\Gamma(f^\prime_2(1525)\to \pi\pi)=2.94$ MeV with the
typical value $R=3.125$ GeV$^{-1}$ is also comparable with the data
$\Gamma_{\pi\pi}=1.4^{+1.0}_{-0.5}$ MeV \cite{Longacre:1986fh}. It
should be noted that the partial width of $f_2^\prime(1525)\to
\pi\pi$ is sensitive to the  mixing angle $\phi$.

\begin{table}[htbp]
\centering%
\caption{Several ratios of the partial decay widths of $f_2(1270)$ and $f_2^\prime(1525)$, and the comparison with the experimental data. \label{BR1}}
\begin{tabular}{cccccccc}
\toprule[1pt]
States&&Ratios&&This work&&Experimental data\\\midrule[1pt]
$f_2(1270)$&&$\Gamma_{K\bar{K}}/\Gamma_{\pi\pi}$&&$0.0239$&& $0.041\pm0.005$ \cite{Nakamura:2010zzi}\\
           &&$\Gamma_{\eta\eta}/\Gamma_{\pi\pi}$&&$0.0073$&& $0.003\pm0.001$ \cite{Barberis:2000cd}\\
           &&$\Gamma_{\eta\eta}/\Gamma_{total}$&&$0.0068$&& $0.004\pm0.0008$ \cite{Nakamura:2010zzi}\\
$f'_2(1525)$&&$\Gamma_{\eta\eta}/\Gamma_{K\bar{K}}$&&$0.0217$&& $0.115\pm0.028$ \cite{Nakamura:2010zzi}\\
           &&$\Gamma_{\pi\pi}/\Gamma_{K\bar{K}}$&&$0.0393$&& $0.075\pm0.035$ \cite{Augustin:1987da}\\
           &&$\Gamma_{\pi\pi}/\Gamma_{\mathrm{total}}$&&$0.0286$&& $0.027^{+0.071}_{-0.013}$  \cite{Gorlich:1979fn}\\
\bottomrule[1pt]
\end{tabular}
\end{table}

Additionally,  several partial decay width ratios of $f_2(1270)$ and
$f_2^\prime(1525)$ are calculated and listed in Table \ref{BR1} to
compare with the corresponding experimental values.

 \subsection{$n^{2S+1}L_J=2^{3}P_{2}$ tensor mesons}\label{n2}

As indicated in Fig. \ref{RT},  the analysis of Regge trajectories
supports $f_2(1640)$ and $f_2(1810)$ to be the candidates of $2
^{3}P_{2}$ tensor states.
Different from these two ground states discussed above,  the
information of the mixing angle of $f_2(1640)$ and $f_2(1810)$ are
still unclear at present. In a realistic picture, $f_2(1640)$ and
$f_2(1810)$ cannot be as pure $n\overline{n}$ and $s\overline{s}$
states, respectively. Thus, we take the range of $\phi$
($\phi=(75\sim 90)^\circ$) when calculating the strong decays of
$f_2(1640)$ and $f_2(1810)$, where $\phi=90^\circ$ denotes
$f_2(1640)/f_2(1810)$ as the pure $n\overline{n}/s\overline{s}$
state.

\begin{center}
\begin{figure}[htb]
\begin{tabular}{c}
\includegraphics[bb=10 175 660 480,scale=0.42,clip]{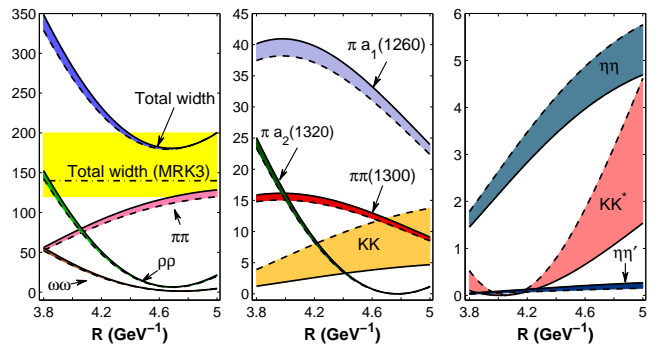}
\end{tabular}
\caption{(color online). The partial decay width of $f_2(1640)$ dependent on the $R$ value
and the comparison with the experimental data (the dash-dot lines with yellow band from Ref. \cite{Bugg:1995jq}). Here, the solid and dashed lines are the results taking the typical values $\phi=90^{\circ}$ and $\phi=75^{\circ}$, respectively. \label{1640}}
\end{figure}
\end{center}

If $f_2(1640)$ are the first radial excitation of $f_2(1270)$,  the
obtained total width of its two-body strong decay overlaps with the
MARK3 data \cite{Bugg:1995jq} as shown in Fig. \ref{1640}. The main
decay channels of $f_2(1640)$ are $\pi\pi$, $\pi a_1(1260)$, and
$\pi\pi(1300)$. In addition, several important decays of $f_2(1640)$
include $\rho\rho$, $\omega\omega$ and $\pi a_2(1320)$, which are
dependent on $R$ value due to the node effect. In experiment,
$f_2(1640)\to \omega\omega$ was reported in Ref. \cite{Alde:1988ea}.
The $4\pi$ channel of $f_2(1640)$ decays was also observed in Ref.
\cite{Alde:1988ea}. If it is due to the $\rho\rho$ contribution, our
calculation turns out to be consistent with this observation. The
$f_2(1640)\to K\bar{K}$ decay width shown in Fig. \ref{1640} is
supported by the experimental observation of Ref.
\cite{Amsler:2006du}. The result in Fig. \ref{1640} indicates that
its $K\bar{K}$, $K\bar{K}^*+h.c.$ and $\eta\eta$ decay channels are
sensitive to the mixing angle $\phi$, which can be interesting
channels to test the mixing angle.

Although we take $f_2(1640)$ as the candidate of the first radial
excitation of $f_2(1270)$,  we can actually compare our result with
the experimental information of $f_2(1565)$ since our results are
insensitive to the mass of the initial state. As listed in PDG
\cite{Nakamura:2010zzi}, the reported decay channel of $f_2(1565)$
are $\pi\pi$, $\rho\rho$, $\eta\eta$, $\omega\omega$, $K\bar{K}$,
which are also supported by our result in Fig. \ref{1640}.

At present, $f_2(1640)$ and $f_2(1565)$ are not classified as the
established states in PDG. According to our calculation, $\pi\pi$ is
the dominant decay channel of $f_2(1640)$. However, the
$f_2(1640)\to \pi\pi$ decay is still missing in experiment. We need
to find a suitable reason to explain the absence of the
$f_2(1640)\to \pi\pi$ decay. Besides the mass difference of
$f_2(1640)$ and $f_2(1565)$, we notice that both $f_2(1640)$ and
$f_2(1565)$ are of similar decay behaviors.
Thus, we suggest further experiment to examine whether $f_2(1640)$
and $f_2(1565)$ are the same state, and clarify why $f_2(1640)\to
\pi\pi$ is absent in experiment while $f_2(1565)\to \pi\pi$ has been
observed. In Ref. \cite{Baker:1999ac}, Baker {\it et al.} once
indicated that $f_2(1565)$ and $f_2(1640)$ could be the same
resonance. Since the $\omega\omega$ decay mode of $f_2(1565)$ has a
relatively high threshold, the resonance peak position is shifted to
higher mass of 1640 MeV.

\begin{center}
\begin{figure}[htb]
\begin{tabular}{c}
\includegraphics[bb=10 175 660 660,scale=0.42,clip]{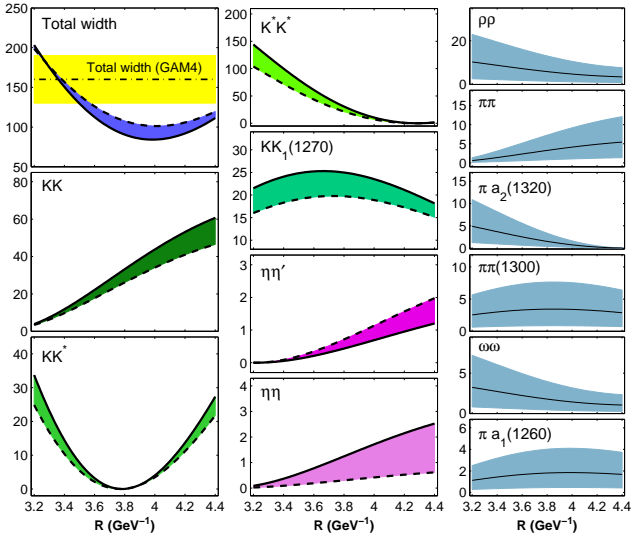}
\end{tabular}
\caption{(color online). The two-body strong decay widths of
$f_2(1810)$ dependent on the $R$ value and mixing angle $\phi$. The
solid and dashed lines in each diagram shown in the first and second
columns correspond to the results taking the typical value
$\phi=85^{\circ}$ and $\phi=75^{\circ}$, respectively. Here, the
experimental width (the dash-dot line with yellow band) of
$f_2(1810)$ is taken from Ref. \cite{Alde:1987rn}. The bands
appearing in the diagrams in the third column are due to
$\phi=(75\sim 85)^\circ$ taken in our calculation, where the solid
lines are the typical values with $\phi=80^\circ$. \label{1810}}
\end{figure}
\end{center}

In Fig. \ref{1810},  the partial decay widths of $f_2(1810)$ as the
first radial excitation of $f_2^\prime(1525)$ are given. We find
that the calculated total width of $f_2(1810)$ is consistent with
the experimental data \cite{Alde:1987rn} when taking $R=3.37$
GeV$^{-1}$. The dominant decay of $f_2(1810)$ is $K^*\bar{K}^*$
while other sizeable decays include $K\bar{K}$, $K\bar{K}^*$,
$KK_1(1270)$. As the mixing state of $n\bar{n}$ and $s\bar{s}$,
$f_2(1810)$ can also decay into $\rho\rho$, $\pi\pi$,
$\pi\pi(1300)$, $\omega\omega$, and $\pi a_2(1320)$, which are shown
in the third column of Fig. \ref{1810}. It shows that the obtained
partial widths are strongly dependent on the mixing angle $\phi$ due
to the dominance of the $s\bar{s}$ component in the $f_2(1810)$
flavor wave function. Taking $\phi=85^\circ/75^\circ$ and $R=3.37$
GeV$^{-1}$, we obtain the typical values of the partial widths of
$f_2(1810)$, i.e. the decay widths of $K^{*}\bar{K}^{*}$,
$K\bar{K}_1(1270)+h.c.$, $K\bar{K}^*+h.c.$,  and $K\overline{K}$ are
respectively $106.2/76.3$ MeV, $23.8/18.0$ MeV, $16.9/12.2$ MeV, and
$10.4/8.6$ MeV.

As the further states listed in PDG \cite{Nakamura:2010zzi}, the
masses of $f_{2}(1750)$ and  $f_2(1810)$ are close to each other.
Because the results presented in Fig. \ref{1810} are not strongly
dependent on the mass of the initial state, we also compare our
result with the experimental data for $f_2(1750)$. It shows that the
calculated  $K\bar{K}$ decay width is consistent with the
experimental data ($\Gamma_{K\bar{K}}=17\pm 5$ MeV) in Ref.
\cite{Shchegelsky:2006et}. Similar to the situation of $f_2(1640)$
and $f_2(1565)$ discussed above, the issue of whether $f_{2}(1810)$
and $f_2(1750)$ can be categorized as the same state should be
clarified in future experiment, especially by the measurement of the
resonance parameters of $f_{2}(1810)$ and $f_2(1750)$.

 \subsection{$n^{2S+1}L_J=3^{3}P_{2}$ tensor mesons}\label{n3}

According to the analysis of mass spectrum in Fig. \ref{RT},
$f_2(1910)$ and $f_2(1950)$ can be candidates of the second radial
excitation of $f_2(1270)$. Since the masses of $f_2(1910)$ and
$f_2(1950)$ are close to each other, it is difficult to distinguish
them only by the mass analysis. We notice the large difference of
the widths of $f_2(1910)$ and $f_2(1950)$. Namely, the average value
of the width of $f_2(1950)$ is $472\pm18$ MeV, which is
significantly larger than that of $f_2(1910)$, i.e.
$\Gamma_{f_2(1910)}=196\pm31$ MeV \cite{Nakamura:2010zzi}. Thus, the
strong decay study of tensor meson with $3 ^{3}P_{2}$ can tell us
which state is suitable to be categorized as the candidate of the
second radial excitation of $f_2(1270)$.

\begin{center}
\begin{figure}[htb]
\begin{tabular}{c}
\includegraphics[bb=30 175 660 680,scale=0.55,clip]{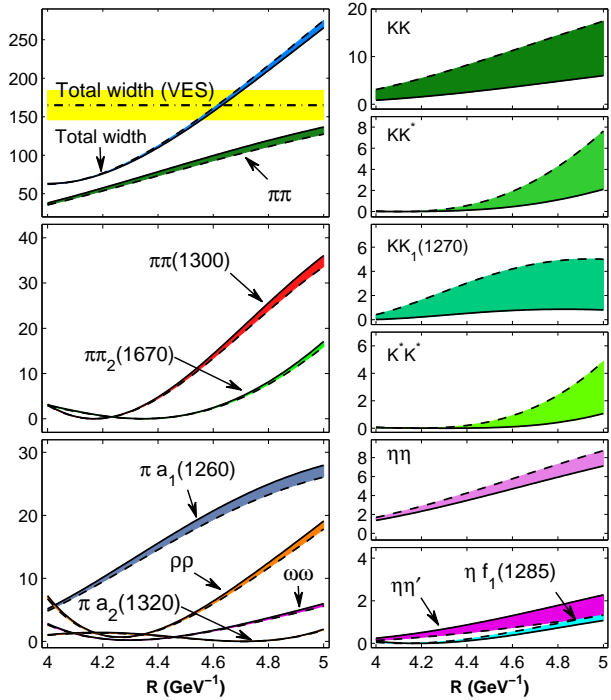}
\end{tabular}
\caption{(color online). The partial decay width of $f_2(1910)$
dependent on the $R$ value and the comparison with the experimental
data (the dash-dot lines with yellow band from Ref.
\cite{Amelin:2006wg}). The partial widths presented here are
arranged in the same way as in Fig. \ref{1640}. \label{1910}}
\end{figure}
\end{center}

As shown in Fig. \ref{1910},  under the $3 ^{3}P_{2}$ assignment to
$f_2(1910)$, the calculated total width of the $f_2(1910)$ two-body
strong decays is in agreement with the experimental width given in
Ref. \cite{Amelin:2006wg}, where we take $R=4.55\sim4.70$
GeV$^{-1}$. Our calculation also provides the information of its
dominant decay ($\pi\pi$) and other sizeable decays ($\pi\pi(1300)$,
$\pi\pi_2(1670)$, $\pi a_1(1260)$, $\rho\rho$, $\omega\omega$, and
$K\bar{K}$). The data for $f_2(1910)\to K\bar{K}$, $\eta\eta$,
$\omega\omega$, $\eta\eta^\prime$ and $\rho\rho$ are available in
experiment \cite{Nakamura:2010zzi}, and we present several ratios
between the partial decay widths to compare with the data:
\begin{eqnarray}
\frac{\Gamma_{\omega\omega}}{\Gamma_{\eta\eta^\prime}}=1.8\sim 2.9,
\quad \frac{\Gamma_{\rho\rho}}{\Gamma_{\omega\omega}}=3.4\sim 3.8
\end{eqnarray}
with $R=4.55\sim4.70$ GeV$^{-1}$. The corresponding measurements are
${\Gamma_{\omega\omega}}/{\Gamma_{\eta\eta^\prime}}=2.6\pm0.6$  and
${\Gamma_{\rho\rho}}/{\Gamma_{\omega\omega}}=2.6\pm0.4$
\cite{Barberis:2000kc}, respectively.

Furthermore, it shows that $f_2(1950)$ is not favored to be
classified as the second radial excitation of $f_2(1270)$ since the
obtained total width in Fig. \ref{1910} are far smaller than the
width of $f_2(1950)$.

\begin{center}
\begin{figure}[htb]
\begin{tabular}{c}
\includegraphics[bb=10 175 670 670,scale=0.42,clip]{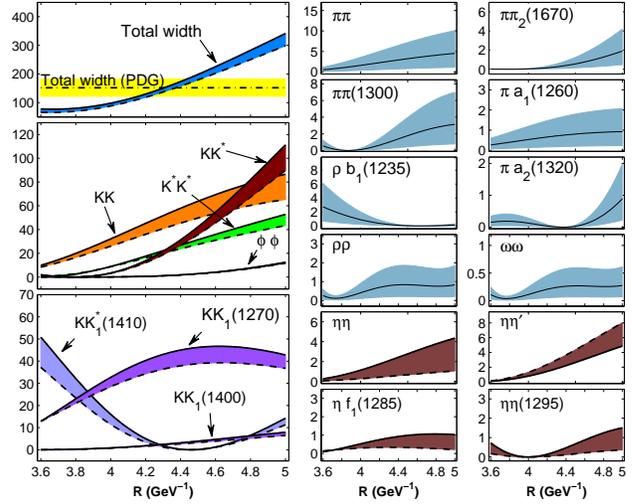}
\end{tabular}
\caption{(color online). The two-body strong decay behavior of the
second radial excitation of $f_2^\prime(1525)$ dependent on the $R$
value and mixing angle $\phi$. The solid and dashed lines in each
diagram s correspond to the results taking the typical value
$\phi=85^{\circ}$ and $\phi=75^{\circ}$, respectively. Here, the
experimental width (the dash-dot line with yellow band) of
$f_2(2150)$ is taken from Ref. \cite{Nakamura:2010zzi}. The bands
appearing in the diagrams corresponding to $\rho b_1(1235)$,
$\pi\pi_2(1670)$, $\pi\pi$, $\pi a_2(1320)$, $\pi\pi(1300)$, $\pi
a_1(1260)$, $\rho\rho$, $\omega\omega$ channels are due to
$\phi=(75\sim 85)^\circ$ taken in our calculation, where the solid
lines are the typical values with $\phi=80^\circ$. \label{2140}}
\end{figure}
\end{center}

In the following analysis, we discuss the second radial excitation
of $f_2^\prime(1525)$.  In Table. \ref{exp}, there are 2 states
$f_2(2140)$ and $f_2(2150)$ with masses near that obtained by the
analysis of Regge trajectories, although $f_2(2140)$ is not a well
established state in PDG \cite{Nakamura:2010zzi}. In our analysis,
we take the mass of the second radial excitation of
$f_2^\prime(1525)$ as $2140$ MeV, and study its decay behavior to
compare with the experimental data of $f_2(2140)$ and $f_2(2150)$.

Sizeable decay modes of the second radial excitation of
$f_2^\prime(1525)$ include $K\bar{K}$, $K\bar{K}^*+h.c.$,
$K^*\bar{K}^*$, and $KK_1(1270)+h.c.$, which suggests the dominance
of strange meson pair decays for $f_2^\prime(1525)$. The calculated
total width supports $f_2(2150)$ as the candidate of the second
radial excitation of $f_2^\prime(1525)$ with $R=4.16\sim4.51$
GeV$^{-1}$. It overlaps with the average width of $f_2(2150)$
($\Gamma_{f_2(2150)}=152\pm30$ MeV) but deviates from that of
$f_2(2140)$ ($\Gamma_{f_2(2140)}=49\pm28$ MeV)
\cite{Nakamura:2010zzi}. The results presented in Fig. \ref{2140}
provide a guidance for further experimental study of the second
radial excitation of $f_2^\prime(1525)$.

 \subsection{$n^{2S+1}L_J=4^{3}P_{2}$ tensor mesons}

\begin{center}
\begin{figure}[htb]
\begin{tabular}{c}
\includegraphics[bb=10 175 660 680,scale=0.43,clip]{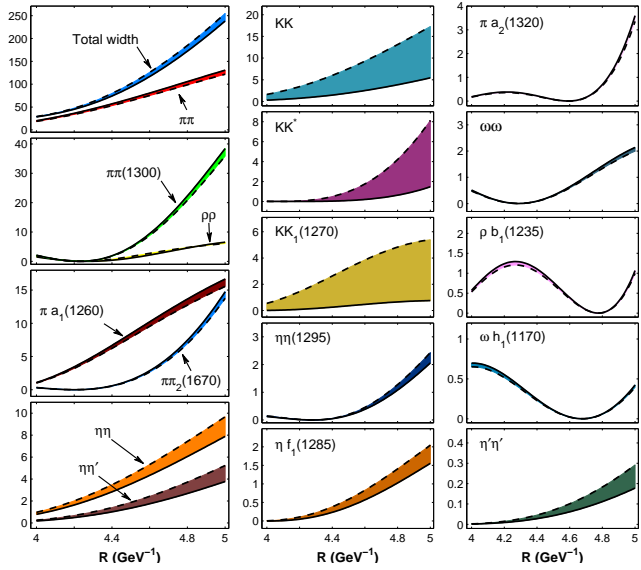}
\end{tabular}
\caption{(color online).  The partial decay width of the third
radial excitation of $f_2(1270)$ dependent on the $R$ value. The
partial widths presented here are arranged in the same way as in
Fig. \ref{1640}. \label{2220}}
\end{figure}
\end{center}

\begin{center}
\begin{figure}[htb]
\begin{tabular}{c}
\includegraphics[bb=10 170 670 670,scale=0.42,clip]{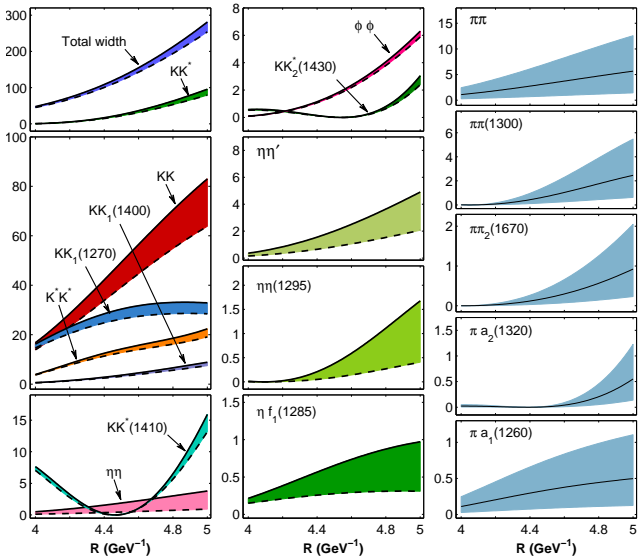}
\end{tabular}
\caption{(color online). The two-body strong decay behavior of the
third radial excitation of $f_2^\prime(1525)$ dependent on the $R$
value and mixing angle $\phi$. The partial widths presented here are
arranged in the same way as in Fig. \ref{1810}. \label{2370}}
\end{figure}
\end{center}

According to the analysis of mass spectrum shown in Fig. \ref{RT},
the masses of the third radial excitations of $f_2(1270)$ and
$f_2^\prime(1525)$ are around 2219 MeV and 2372 MeV, respectively.
It thus makes $f_2(2240)$ and $f_2(2410)$ good candidates for the
third radial excitations. The calculations of their partial decay
widths and total widths of two-body strong decays are presented in
Figs. \ref{2220} and \ref{2370}.

From Fig. \ref{2220}, it shows that the decay of $f_2(2240)$ is
dominated by the $\pi\pi$ and $\pi\pi(1300)$ channels. Other
sizeable decay channels include $\rho\rho$, $\pi\pi_2(1670)$, $\pi
a_1(1260)$, $\eta\eta$, and $\eta\eta^\prime$ etc. As the candidate
of the third radial excitation of $f_2^\prime(1525)$, $f_2(2410)$
mainly decays into strange meson pairs due to the dominance of the
$s\bar{s}$ component in its wavefunction. As shown in Fig.
\ref{2370}, the $f_2(2410)\to K\bar{K}$ is dominant while other
decay channels such as $K\bar{K}^*+h.c.$, $K\bar{K}_1(1270)+h.c.$,
$K\bar{K}_1(1400)$, and $K^*\bar{K}^*$ are also important.

Although $f_2(2240)$ and $f_2(2410)$ can be taken as good candidates
for the third radial excitations of $f_2(1270)$ and
$f_2^\prime(1525)$, respectively, detailed experimental information
is still absent for further understanding of their properties.
It should be useful to compare the calculated decay widths shown in
Figs. \ref{2220} and \ref{2370} with future experimental
measurements in order to gain further insights into the nature of
these two states.

 \subsection{$n^{2S+1}L_J=1^{3}F_{2}$ tensor mesons}

\begin{center}
\begin{figure}[htb]
\begin{tabular}{c}
\includegraphics[bb=10 170 660 680,scale=0.46,clip]{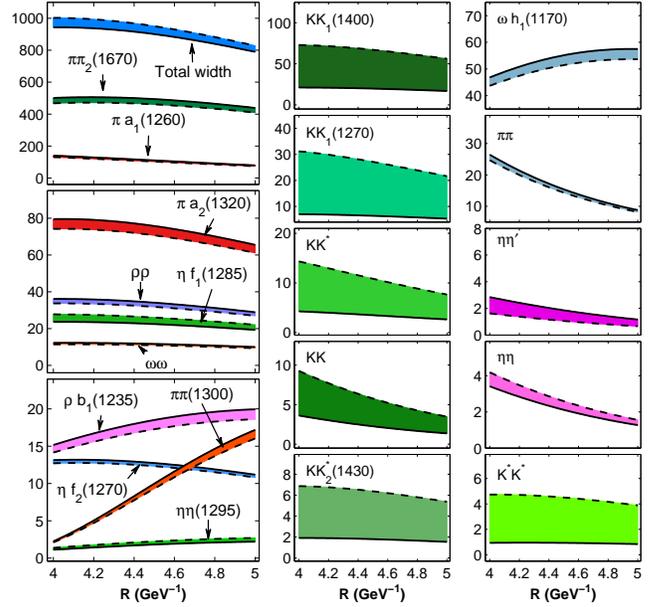}
\end{tabular}
\caption{(color online).  The partial decay width of tensor meson
with $1^{3}F_{2}$ quantum number dependent on the $R$ value. The
partial widths presented here are arranged in the same way as in
Fig. \ref{1640}. \label{2010}}
\end{figure}
\end{center}

In Fig. \ref{2010}, we present the partial decay widths of the
ground state  $F$-wave tensor meson with $1^{3}F_{2}$. Assuming that
it is dominated by the $n\bar{n}$ component, we use the mass of
$f_2(2010)$ as the input to calculate the partial decays widths. It
shows that the width of the $1^{3}F_{2}$ tensor meson is very broad.
Some of those dominant decay channels can be seen in Fig. \ref{2010}
in terms of $R$, e.g. $f_2(2010)\to \pi\pi_2(1670)$ and $\pi
a_2(1260)$.

\begin{center}
\begin{figure}[htb]
\begin{tabular}{c}
\includegraphics[bb=10 170 630 670,scale=0.43,clip]{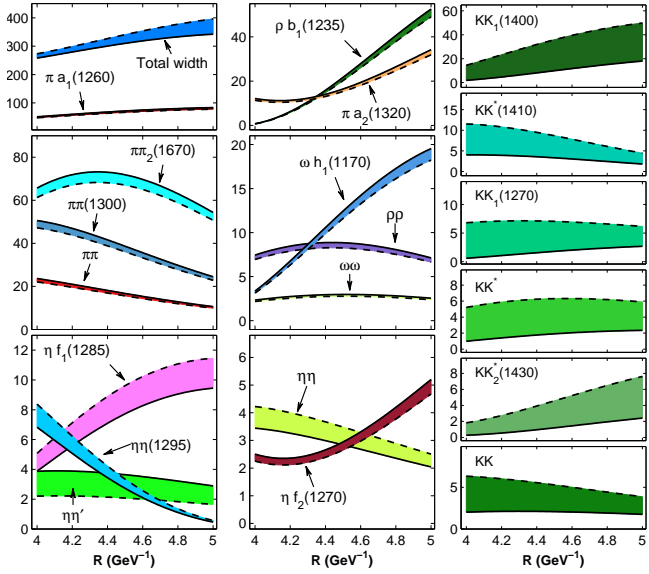}
\end{tabular}
\caption{(color online). The partial decay width of tensor meson
with  $2^{3}F_{2}$ quantum number dependent on the $R$ value. The
partial widths presented here are arranged in the same way as in
Fig. \ref{1640}. \label{2300}}
\end{figure}
\end{center}

In this work, we also study the two-body decays of the $2^{3}F_{2}$
tensor meson,  which is the second radial excitation of
$1^{3}F_{2}$. With the mass of $f_2(2300)$ as the input, the
calculated partial decay widths are shown in Fig. \ref{2300}. A sum
of these two-body decay partial widths also gives a broad total
width for this state.

At present, the experimental information for $f_2(2010)$,
$f_2(2300)$ and $f_2(2340)$ is not enough to help us establish them
as the candidates of $1^{3}F_{2}$ and $2^{3}F_{2}$ tensor mesons. We
also expect more experimental measurements in the future would be
able to clarify their properties.

\section{Discussion and conclusion}\label{sec5}

So far there have been a large number of isoscalar tensor states
observed in experiment \cite{Nakamura:2010zzi}.  A crucial task is
to understand their properties and try to categorize them into
hadron spectrum. In this work we carry out a systematic study of
these isoscalar tensor states by performing an analysis of the mass
spectrum and calculating their two-body strong decays in the QCP
model. By comparing our results with the available experimental
data, we extract important information for a better understanding of
the underlying structures of these $f_2$ states.

By the analysis of mass spectrum based on the Regge trajectories of
tensor states, we identify candidates for the radial excitations of
the $P$-wave and $F$-wave states. By calculating the partial decay
widths of those states into various two-body final state hadrons and
comparing the results with the available experimental information,
we succeed in categorizing some of those states into the $q\bar{q}$
radial excitation spectrum. Meanwhile, some obvious controversies
are also exposed, for which further theoretical and experimental
studies are needed. In particular, at present, only
$f_2(1270)/f_2^\prime(1525)$ have a relatively well-established
experimental status, while experimental information for other
observed higher $f_2$ states is still limited.
Although $f_J(2220)$ has been proposed to be a tensor glueball
candidate in the literature \cite{Giacosa:2005bw,Burakovsky:2000gk},
further study of the tensor glueball is still needed in order to
single it out from experimental observables. Improved experimental
measurements of those higher $f_2$ states would be crucial for
distinguish the conventional $q\bar{q}$ scenario from other possible
exotic configurations.

\vfil

\section*{Acknowledgements}

This project is supported in part by the National Natural Science
Foundation of China (Grants No. 11175073, No. 11035006),  the
Ministry of Education of China (FANEDD under Grants No. 200924,
DPFIHE under Grants No. 20090211120029, NCET under Grants No.
NCET-10-0442, the Fundamental Research Funds for the Central
Universities), the Fok Ying-Tong Education Foundation (No. 131006)
and the Chinese Academy of Sciences (KJCX2-EW-N01).

\end{document}